\documentclass[%
 reprint,
superscriptaddress,
 amsmath,amssymb,
 aps,
pre,
]{revtex4-2}
\usepackage{CJKutf8}
\usepackage[english]{babel} 
\usepackage[pdftex]{graphicx}
\usepackage{placeins}
\usepackage{dcolumn}
\usepackage{bm}
\usepackage{url} 
\usepackage{xcolor}
\usepackage{subfigure}
\usepackage{comment}
\addto\captionsenglish{}

\usepackage{titlesec}
\titlespacing*{\section}
{0pt}{6ex}{1ex}
\titlespacing*{\subsection}
{0pt}{6ex}{1ex}

\begin{document}
\begin{CJK*}{UTF8}{min}


\title{Percolation without trapping: how Ostwald ripening during two-phase displacement in porous media alters capillary pressure and relative permeability}

\author{Ademola Isaac Adebimpe}
    \email[Corresponding author: ]{a.adebimpe21@imperial.ac.uk}
    \affiliation{Department of Earth Science and Engineering, Imperial College London, London, SW7 2BP, UK.}
    \affiliation{Department of Chemical Engineering, Obafemi Awolowo University, Nigeria.}
\author{Sajjad Foroughi}%
\author{Branko Bijeljic}%
\author{Martin J. Blunt}%
\affiliation{Department of Earth Science and Engineering, Imperial College London, London, SW7 2BP, UK.}%
\date{June 14, 2024}
\begin{abstract}
Conventional measurements of two-phase flow in porous media often use completely immiscible fluids, or are performed over time-scales of days to weeks.  If applied to the study of gas storage and recovery, these measurements do not properly account for Ostwald ripening, significantly over-estimating the amount of trapping and hysteresis.  When there is transport of dissolved species in the aqueous phase, local capillary equilibrium is achieved: this may take weeks to months on the centimetre-sized samples on which measurements are performed.  However, in most subsurface applications where the two phases reside for many years, equilibrium can be achieved.  We demonstrate that in this case, two-phase displacement in porous media needs to be modelled as percolation without trapping.  A pore network model is used to quantify how to convert measurements made ignoring Ostwald ripening to correct trapped saturation, capillary pressure and relative permeability to account for this effect.  We show that conventional measurements over-estimate the amount of capillary trapping by 20-25\%. 
\end{abstract} 

\maketitle
\clearpage\end{CJK*}

\section{INTRODUCTION}
Since the seminal work of Wilkinson \& Willemnsen  \cite{Wilkinson1983} and Lernomand {\it et al.} \cite{Lenormand1983}, immiscible multiphase flow in porous media has been described as a percolation-like process, including trapping when one phase is completely surrounded by another.  A trapped cluster is assumed to retain its volume and capillary pressure at which it was first immobilized.  However, in many natural and engineered settings, including subsurface gas storage and recovery, water table movement and transport through gas diffusion layers in electrolysers and fuel cells, there is also transport of dissolved components leading to Ostwald ripening \cite{ostwald1897studien}.  Differences in local capillary pressure lead to small differences in solubility, establishing concentration gradients in the aqueous phase.  By diffusion, gas dissolved in water is transported between ganglia, allowing for rearrangement of the trapped phase.  In equilibrium, the concentration of dissolved gas is constant everywhere which means that there is local capillary equilibrium, even for trapped clusters of the gaseous phase.

Most experimental measurements of multiphase flow properties are conducted on rock samples a few cm in size using either completely immiscible fluids (such as oil and water, or mercury and a vacuum), or are conducted over a period of days to a few weeks, during which time equilibrium of capillary pressure may not be achieved. On the other hand, in most subsurface situations, the two phases will reside with little or no flow for periods of months or years. As we demonstrate later, in equilibrium of both capillary forces and dissolved concentration, the correct model of displacement is percolation {\it without} trapping.  Nonetheless, phases can still be trapped.  We will show how to correct the residual (trapped) saturations, capillary pressures and relative permeabilities, including the effect of hysteresis, to account for Ostwald ripening.

Consider two fluid phases in the pore space, a denser phase, labelled 1, and a less dense phase, 2.  The capillary pressure, $P_c = P_2 - P_1$, is imposed by setting the pressures in each phase: for infinitesimal flow rates, where the phases are connected to the inlet or outlet, the capillary pressure is constant.  The pore space can be conceptualized as a network of wide regions, pores, connected by restrictions, called throats \cite{Blunt2017}.

Let us assume that phase 1 is wetting and phase 2 non-wetting. Drainage, the displacement of a wetting phase by a non-wetting phase during which the capillary pressure increases, is invasion percolation, where the non-wetting phase fills the largest throat that is available for filling through a connected pathway of non-wetting phase from the inlet \cite{Wilkinson1983}.  The adjacent pore, being larger than the throat, is also filled. If it is assumed that there are connected wetting layer in roughness and corners in the pore space, the wetting phase is not trapped.  The reverse process, imbibition, is displacement of the non-wetting phase by a wetting phase as the capillary pressure decreases.  This is an ordinary percolation process since the wetting phase has access to the whole pore space through wetting layers \cite{Lenormand1984}: the wetting phase occupies throats in order of size, from the smallest upwards, by snap-off \cite{Pickell1966}. However, once the non-wetting phase is completely surrounded by the wetting phase, it is trapped, and hence imbibition is percolation with trapping.  When a cluster of the non-wetting phase is trapped, it assumes a local capillary pressure that may be different from the prevailing capillary pressure imposed on the continuous phases \cite{Andrew2015}. 

While there are additional subtleties, including complex filling dynamics and Haines jumps in drainage \cite{Haines1930, Berg2013}, intermittency \cite{Reynolds2017}, and cooperative pore filling in imbibition \cite{Lenormand1983, Lenormand1984}, this model of quasi-static displacement has remained unchallenged for 40 years.  With an accurate representation of the pore-space geometry, and including the percolation-like processes mentioned above in a pore network model, qualitative predictions of the multiphase flow properties, capillary pressure and relative permeability, can be made (see, for instance, \cite{Oren1998, Valvatne2004}).

\section{OSTWALD RIPENING}
If the two fluids are not completely immiscible, for instance if phase 2 is a gas (air, natural gas, carbon dioxide or hydrogen), displacement is also accompanied by mass transport of dissolved species through phase 1 (usually water, but can be oil in the presence of carbon dioxide or natural gas).  This facilitates Ostwald ripening, where diffusive transport in the aqueous phase acts to make the local capillary pressure uniform \cite{Xu2017, Garing2017,deChalendar2018, Singh2022}.  Ostwald ripening in porous media has been studied experimentally, theoretically and numerically: it leads to rearrangement of the phases in the pore space with an overall trend of increasing connectivity \cite{zhang2023pore, goodarzi2024trapping}.

\subsection{Equilibrium model for partially miscible two-phase flow}
If we consider slow displacement, what is the correct equilibrium model if Ostwald ripening leads to a constant capillary pressure, even for disconnected phases?  Ostwald ripening ensures that locally the capillary pressure is uniform, even in disconnected clusters.  Consider the displacement of phase 2 by phase 1  where the capillary pressure decreases. A cluster of phase 2 is first trapped when it is completely surrounded by phase 1, and its capillary pressure is the value imposed on the connected phases. As displacement proceeds, and the capillary pressure drops, the capillary pressure in the trapped cluster is now higher than in the connected phases.  If we allow Ostwald ripening, this means, from Henry's law, that the concentration of dissolved gas adjacent to the cluster is slightly higher than next to the connected phase, creating a concentration gradient. If we allow sufficient time for mass transport by diffusion, material from the trapped cluster will dissolve and be transferred to the connected phase 2.  

Now imagine a displacement event possible at the prevailing capillary pressure, but which occurs in a trapped cluster that would be forbidden in a conventional model with trapping.  Ostwald ripening will mean that removal of gas is possible, albeit through diffusion through phase 1, rather than by immiscible displacement. In the end this is simply a percolation-like model where trapping is ignored. Phase 2 does remain trapped at the end of the displacement.  Displacement stops when we can no longer impose a capillary pressure through connected phases: this occurs when all of phase 2 is first disconnected.

The same argument pertains for subsequent injection of phase 2, allowing for the reconnection of previously-trapped clusters.  Displacement is possible, without trapping of phase 1, from any cluster containing phase 2, even if it is disconnected.  We assume, however, that Ostwald ripening does not allow clusters of phase 2 to nucleate in parts of the pore space occupied by phase 1.  When phase 1 is wetting, this has a less significant impact on the behaviour, as phase 1 is not trapped anyway, thanks to layer flow; however, unlike a traditional invasion percolation model, phase 2 can displace phase 1 from a disconnected cluster.

\subsection{Time-scales for equilibrium and experimental implications}
Having established that the equilibrium model of displacement is percolation without trapping, whereas it is percolation with trapping for completely immiscible fluids, we need to explore the implications for the analysis of field-scale gas storage and recovery based on experimental measurements, typically made on samples a few centimetres in size.

The first step is to consider if there is sufficient time for mass transport through the aqueous phase to reach equilibrium.  This has previously been analysed analytically \cite{blunt2022ostwald}.
A typical timescale for equilibrium established by diffusion, $t_d$, over a length $l$ is \cite{blunt2022ostwald}
   	\begin{equation}
	\label{Eq:time}
	t_d = \frac {l^2 r \phi S_{gr}\rho_g}  {2 D H \sigma m_g },
\end{equation}	
\noindent where $r$ is a typical throat radius, $m_g$ is the molecular mass of the gas, $S_{gr}$ is the residual gas saturation, $\phi$ is the porosity, $\sigma$ is the gas-water interfacial tension, $\rho_g$ is the gas density, $D$ is the diffusion coefficient of dissolved gas in the aqueous phase and $H$ is the Henry's constant. Eq.~(\ref{Eq:time}) finds the time sufficient to transport an amount of gas equivalent to the residual saturation $S_{gr}$ entirely through the aqueous phase by molecular diffusion.

However, the pore-space arrangement of fluids is also determined by advection, or the flow of connected gas in its own phase under imposed or natural pressure gradients.  If we assume a Darcy flux (volume per unit area per unit time) $q$, then for a cubic volume of side $l$, the volume of gas transported in a time $t_a$ is $l^2 q t_a$.  If we equate this to the volume of residual gas in the same volume $l^3 \phi S_{gr}$, we can define an advective timescale:
   	\begin{equation}
	\label{Eq:advection}
	t_a = \frac {l \phi S_{gr}}{q}.
\end{equation}
The ratio of diffusive and advective times is therefore
   	\begin{equation}
	\label{Eq:ratio}
	R = \frac {t_d}{t_a} = \frac {l q r \rho_g} {2 D H \sigma m_g}.
\end{equation}

As an example, using typical values of the parameters in Eq.~(\ref{Eq:time}) (see Table 1 in \cite{blunt2022ostwald} for details), the time to reach equilibrium over a length $1 = 10^{-3}$ m (taken as a representative distance between trapped ganglia of gas) by diffusion at subsurface conditions is $1.5 \times 10^6$ s, $3 \times 10^6$s and $9\times 10^6$~s for carbon dioxide, hydrogen and methane respectively: this is approximately 17 to 100 days. In comparison, advection is much more rapid, assuming that the gas is connected through the pore space: taking a flow rate of $10^{-7}$~ m/s, $\phi = 0.2$ and $S_{gr}= 0.3$, $t_a$ is only 600~s in Eq.~(\ref{Eq:advection}).

From Eq.~(\ref{Eq:ratio}) it is evident that as we consider larger length-scales, $l$, from centimetre-sized samples used for experiments to the kilometre extent of gas storage, advection becomes increasingly dominant.  Hence we can assume that when there is flow of gas -- it is not completely trapped -- advection will control how it moves and its occupancy in the pore space for all reasonable values of diffusion coefficient, Henry's constant and interfacial tension.

The conceptual picture of flow in the presence of Ostwald ripening is therefore as follows:  Advection will dominate the pore-scale arrangement of fluid when the gas is well-connected in the pore space.  While there can be transport of dissolved material, since the connected phase maintains a constant capillary pressure locally, there are no significant concentration gradients or transport of dissolved gas in the aqueous phase.  The situation when a significant fraction of the gas is trapped -- at or near the end-point of imbibition or the start of secondary drainage -- requires more careful consideration, however.  Here there can be a difference in capillary pressure between trapped ganglia and the connected phase, as discussed in the previous section. Advective flow is controlled by the multiphase flow functions, relative permeability and capillary pressure, that are determined by the pore-scale arrangement of fluid.  These functions do change with Ostwald ripening, as the distribution of phases within the pore space alters; hence, even if advection is fast, Ostwald ripening has an effect if it alters the multiphase flow functions during displacement.  The question is whether there is sufficient time for Ostwald ripening to have a significant impact on relative permeability and capillary pressure.

In experimental studies, a full cycle of injection on a centimetre-sized sample is normally completed in a few days to weeks, dependent on the methodology used \cite{Blunt2017}.  What is of principal concern here is the amount of time, near the end of the displacement, for rearrangement of trapped phases.  In this case, it is unlikely that equilibrium is reached at the mm-scale such that there is significant reconnection and flow, and certainly not at the scale of the whole sample.  On the other hand, in gas storage operations, after injection the system is left for typically 3-6 months before withdrawal (for hydrogen and natural gas), or, in the case of carbon dioxide sequestration, the supercritical gas is left to migrate through the formation under natural pressure gradients and buoyancy over decades to millennia.  In these circumstances it is likely that equilibrium is achieved at a scale sufficient for phases potentially to reconnect and flow.  It is not necessary for equilibrium to be satisfied by diffusion alone over much longer lengths, since once the gas reconnects, advection leads to a much more rapid transport.

In this work we hypothesize that under most experimental conditions, Ostwald ripening has an insignificant effect on the measured multiphase flow properties such that displacement can be modelled as percolation with trapping.  Under field conditions we assume, in contrast, that Ostwald ripening leads to capillary pressure equilibrium, even for trapped clusters, and can be modelled as percolation without trapping.  We will now use a pore network model to explore what impact this has on capillary pressure, relative permeability and residual saturation.

\section{METHODOLOGY}
We now explore the differences between a percolation model with and without trapping on predictions of residual saturation, capillary pressure and relative permeability using a pore network model that can simulate an arbitrary sequence of displacement both with trapping (the conventional model) and without (considering Ostwald ripening).  The input is an unstructured network of pores connected by throats extracted from pore-space images by the maximal ball method \cite{Dong2009, Raeini2017}.  The model, with trapping, is identical in terms of algorithm to previous work \cite{Valvatne2004}.  

We will only consider two extreme cases: no Ostwald ripening (rapid experiments or immisicble fluids), and equilibrium considering Ostwald ripening (which is representative of most subsurface flows).  We will not consider a transient, time-dependent process where both advection and diffusive transport compete.  This has been simulated in pore-scale models by several authors using level set method \cite{singh2022level},  level set with mass transfer method \cite{singh2023pore, singh2024ostwald}, and pore network modelling including pressure depletion \cite{moghadasi2023pore}, and bubble equilibration using sequential \cite{deChalendar2018} and fully-implicit \cite{mehmani2022pore} algorithms.  The purpose of our work is not to propose a new, sophisticated model of the time evolution of phases with Ostwald ripening, but to propose a simple approach to correct experimental results, where the effect is small, to field settings where we are at, or close to, equilibrium.

As a base case we simulated displacement through a network derived from a micro-tomography image of Bentheimer sandstone \cite{bentheimer} that has been widely used as a benchmark for many experimental and modelling studies \cite{Lin2018}.  To explore the sensitivity of our results to pore structure, we also studied networks derived from two images of Estaillades limestone \cite{estaillades_1, estaillades_2}, Ketton, an oolithic limestone \cite{scanziani2020dynamics}, and a reservoir carbonate sample \cite{alhosani2020dynamics}.  In all cases the receding contact angle (measured through phase 1) was zero for drainage and $45.6^{\circ} \pm 20.1^{\circ}$ distributed uniformly for imbibition.  This value was chosen based on measurements of contact angle during displacement and previous modelling studies \cite{Blunt2019, Lin2018, Raeini2018}.

\section{RESULTS AND DISCUSSION}
\subsection{Capillary pressure}
\begin{figure}[t]
    \centering
    \begin{subfigure}{
        \includegraphics[width=0.39\textwidth]{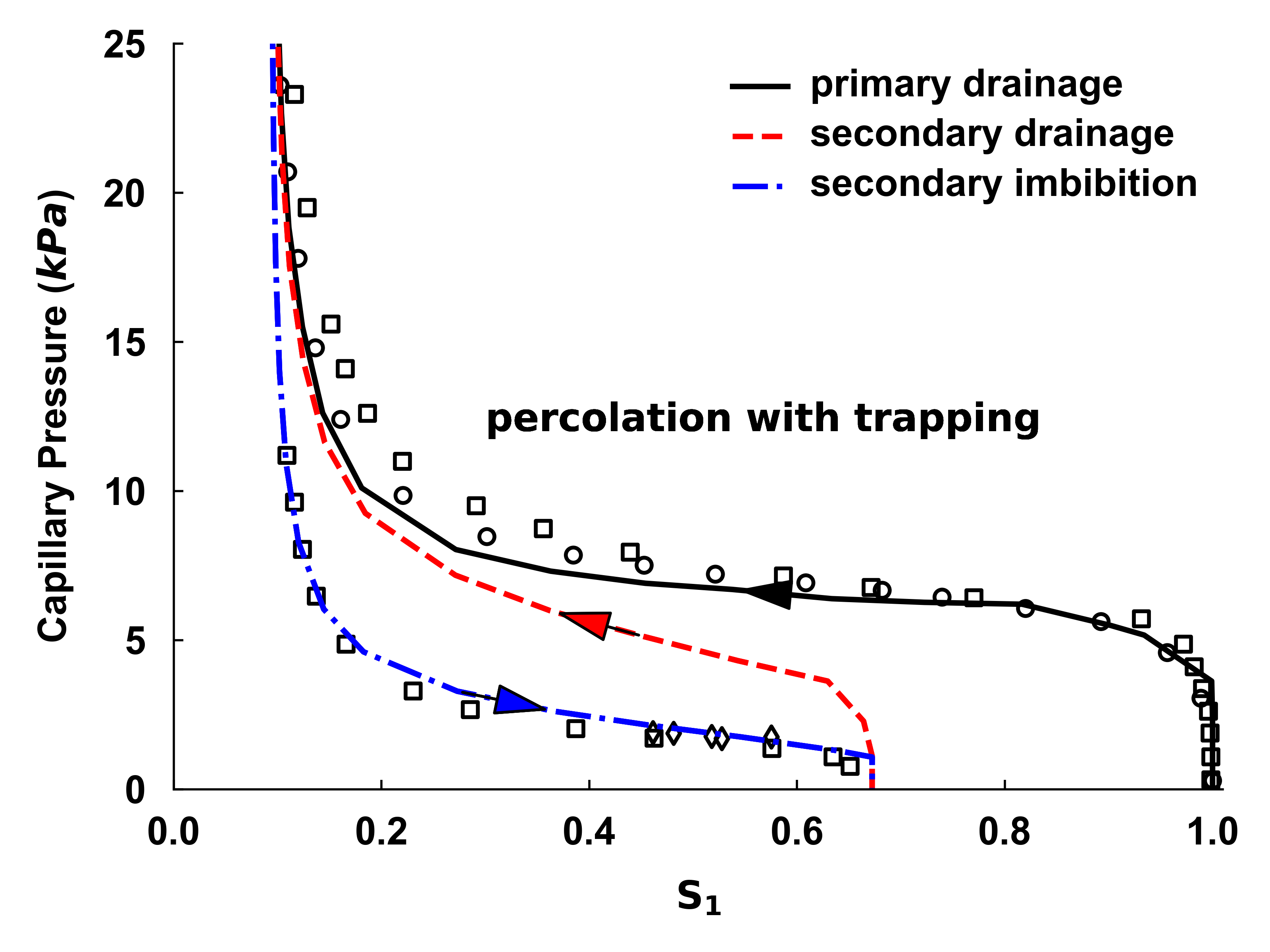}
        \label{PcS1_wt}
        }
    \end{subfigure}
    \begin{subfigure}{
        \includegraphics[width=0.39\textwidth]{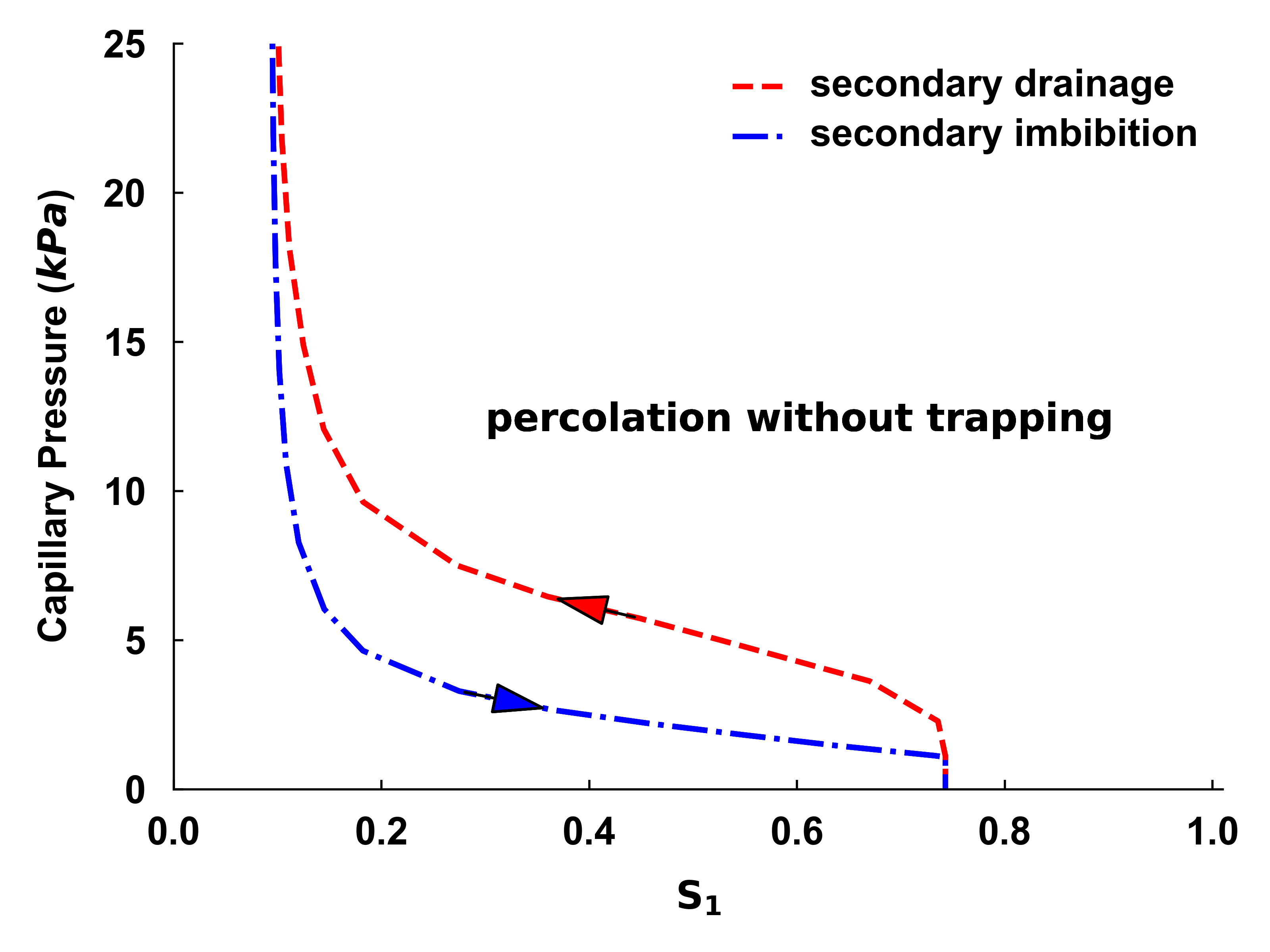}
        \label{PcS1_nt}
        }
    \end{subfigure}
    
    \caption{Capillary pressure as a function of the saturation of phase 1 assuming (a)  filling of a pore or throat is not allowed if trapped, and (b) when filling of a pore or throat is allowed even if trapped.  The imposed capillary pressures in drainage reached  $100$~kPa while during imbibition filling was continued until the non-wetting phase became disconnected.  The lines are model predictions while the points are experimental measurements where Ostwald ripening had not reached equilibrium (percolation with trapping): squares for primary drainage and secondary imbibition from \cite{Raeesi2014}; and circles and diamonds for mercury injection capillary pressure drainage and secondary imbibition respectively from \cite{Lin2018}.}
    \label{PcSw}
\end{figure}

Fig.~\ref{PcSw} shows the predicted capillary pressure for primary drainage, secondary imbibition and secondary drainage with and without trapping in imbibition.  For the case with trapping, the model predictions are compared to experimental measurements of primary drainage using mercury injection \cite{Lin2018}, where Ostwald ripening does not occur, and nitrogen/water experiments for both primary drainage and imbibition \cite{Raeesi2014} where the entire imbibition cycle was performed in 5 to 48 hours.  Based on the analysis in the previous section, we suggest that this is insufficient time for Ostwald ripening to reach equilibrium.  There is a good agreement between the model predictions and experiment, including the trapped saturation $S_{2t}$.

Fig.~\ref{PcSw}(b) presents model predictions when Ostwald ripening is considered.  Here we see a decrease in trapped saturation from approximately 0.36 to 0.28 (see Table~1): the capillary pressure for secondary displacement cycles appears stretched along the saturation axis compared to the case with trapping.

\begin{figure}[t]
    \centering
    \begin{subfigure}{
        \includegraphics[width=0.39\textwidth]{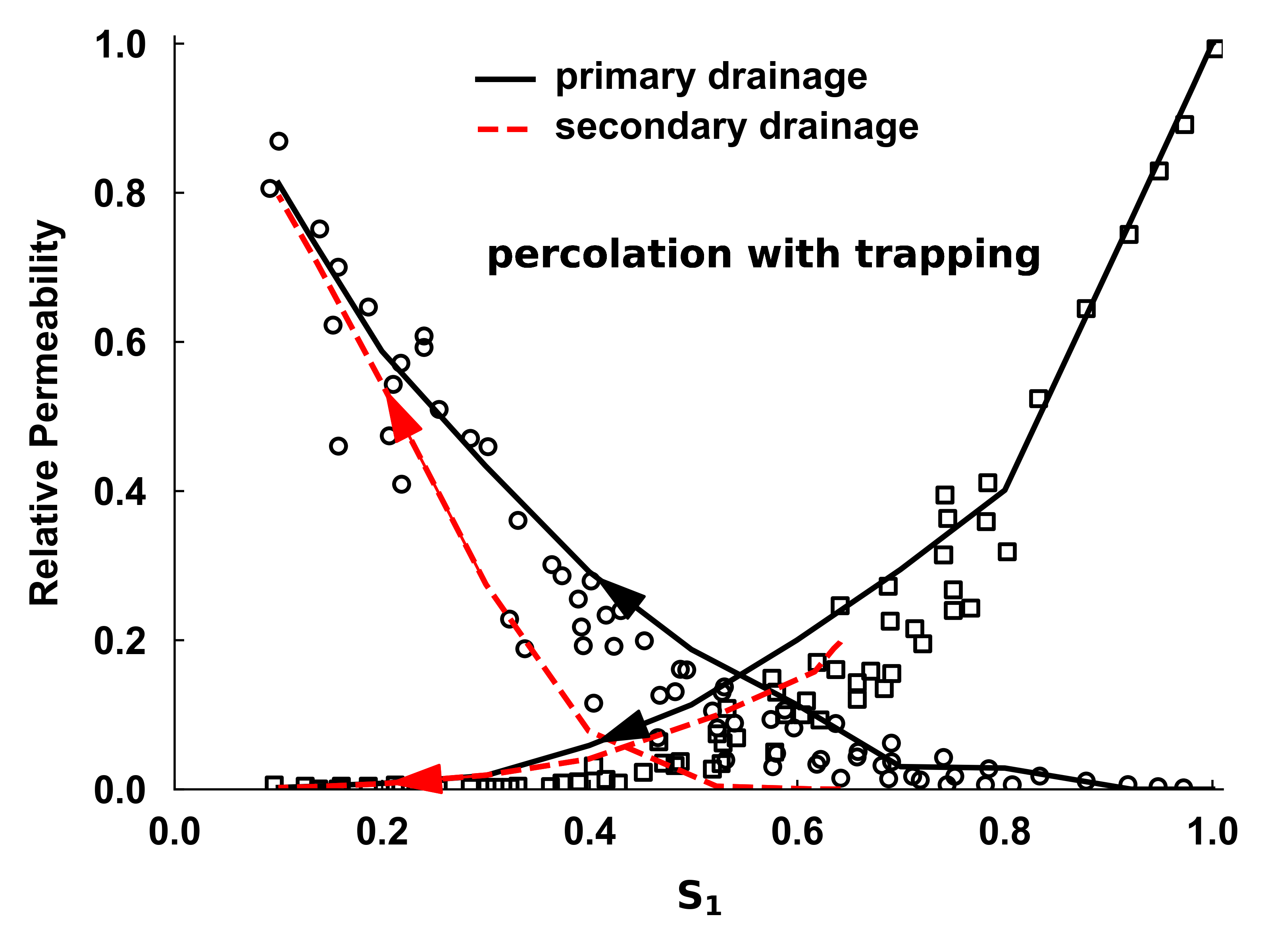}
        \label{krS1_D_wt}
        }
    \end{subfigure}
    \begin{subfigure}{
        \includegraphics[width=0.39\textwidth]{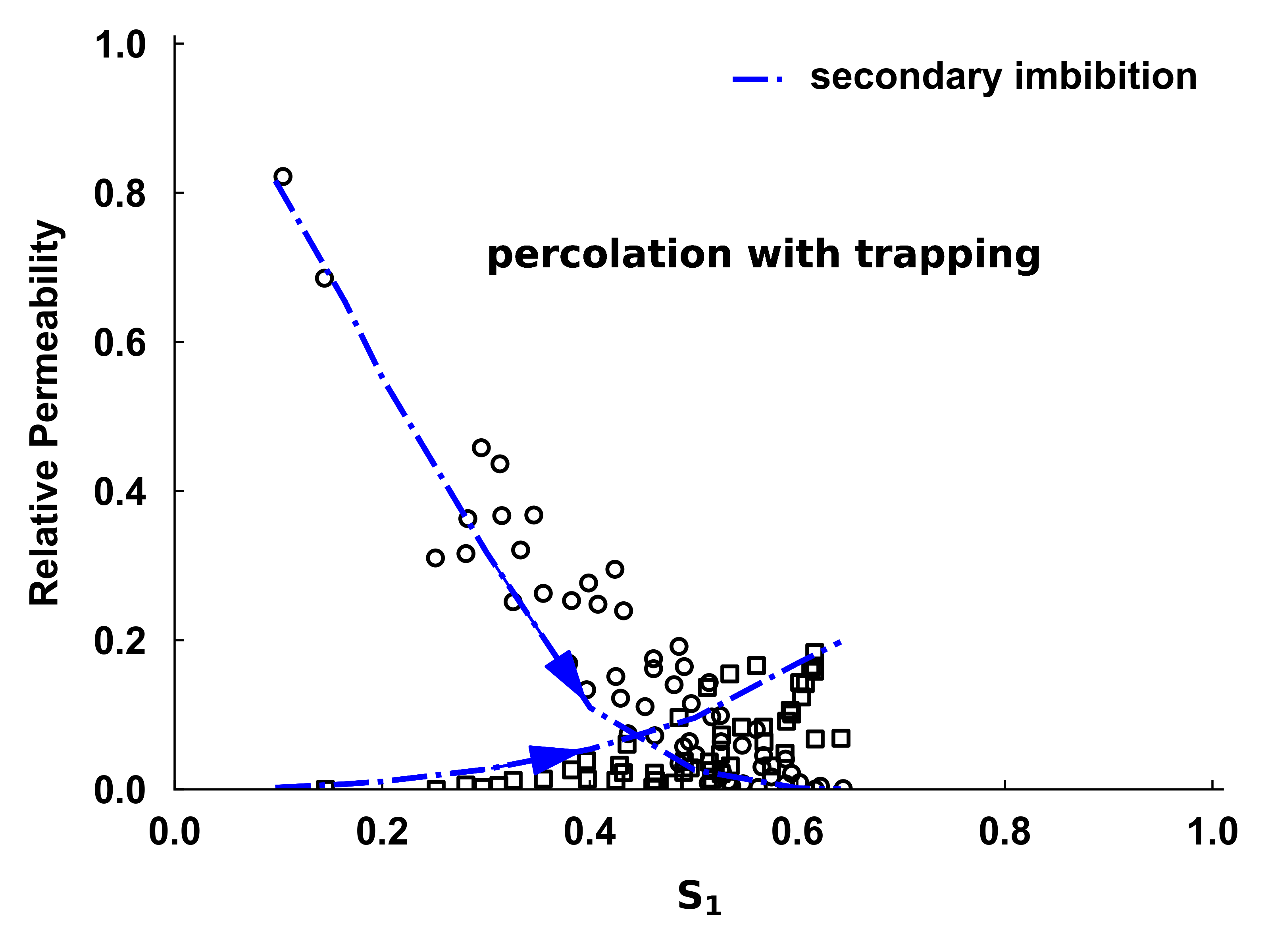}
        \label{krS1_I_wt}
        }
    \end{subfigure}
    \begin{subfigure}{
        \includegraphics[width=0.39\textwidth]{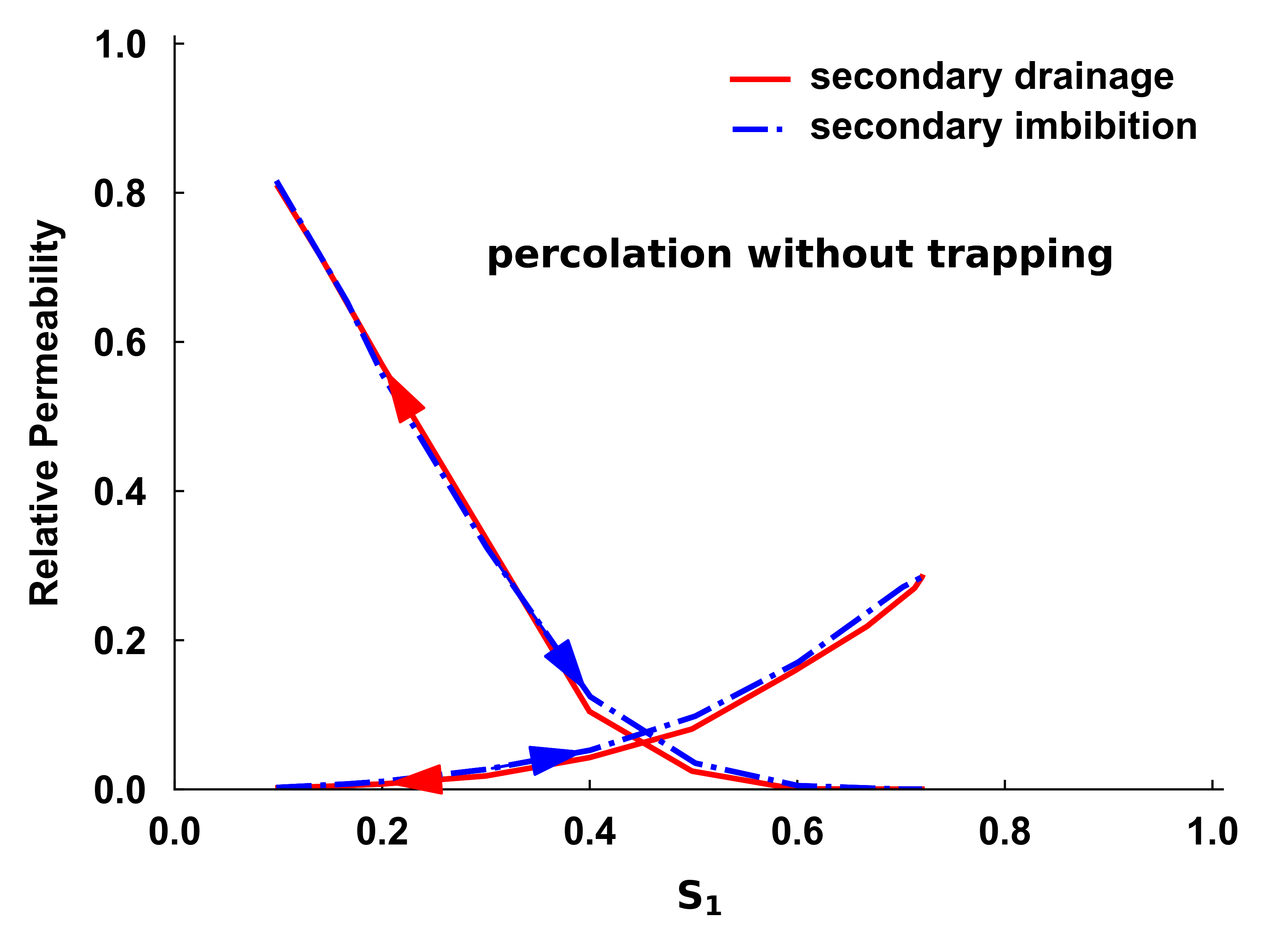}
        \label{krS1_nt}
        }
    \end{subfigure}
    
    \caption{Relative permeability as a function of the saturation of phase 1 for (a) primary and secondary drainage and (b) imbibition assuming filling is not allowed if trapped; and (c) drainage and imbibition assuming filling is allowed even if trapped.  The lines are model predictions while the points are measurements where Ostwald ripening is negligible (percolation with trapping): squares for phase 1 and circles for phase 2 from \cite{Oren1998, Alizadeh2014}.}
    \label{krSw}.
\end{figure}

\subsection{Relative permeability}
Predictions of relative permeability are compared with experimental data \cite{Oren1998, Alizadeh2014} in Fig. \ref{krSw}: here the measurements were made with immiscible oil/water systems where the impact of Ostwald ripening is negligible. The model predictions fall within the considerable scatter of the data, consistent with other work using algorithmically similar network models \cite{Valvatne2004, Raeini2018}.  However, when Ostwald ripening is considered, although the non-wetting phase relative permeability is similar, it reaches a lower residual saturation, while the wetting phase relative permeability reaches a higher maximum value, following the trend in primary drainage: the wetting phase occupies the smallest regions of the pore space, as well as corners and roughness, and reaches a higher conductance as the saturation increases.

\subsection{Proposed scaling to account for Ostwald ripening}
As explored above, most experimental measurements of capillary pressure and relative permeability are not conducted at equilibrium, in terms of Ostwald ripening, and hence are not representative of most subsurface flows involving gases.  This is either because mercury or oil/water systems are used, or the experiment is performed too rapidly to allow diffusive transport to rearrange the gaseous phase.  We now propose how to convert measurements made out of equilibrium (with trapping) to account for Ostwald ripening.

We define a normalized saturation $1 \geq S_e \geq 0$:
\begin{equation}
\label{Eq:Se}
    S_e = \frac{S_1 - S_{1i}}{1-S_{1i}-S_{2t}},   
\end{equation}
where $S_{1i}$ is the initial (irreducible) saturation of phase 1 at the end of primary drainage and $S_{2t}$ is the trapped saturation of phase 2.

\begin{figure}[!]
    \centering
    \begin{subfigure}{
        \includegraphics[width=0.39\textwidth]{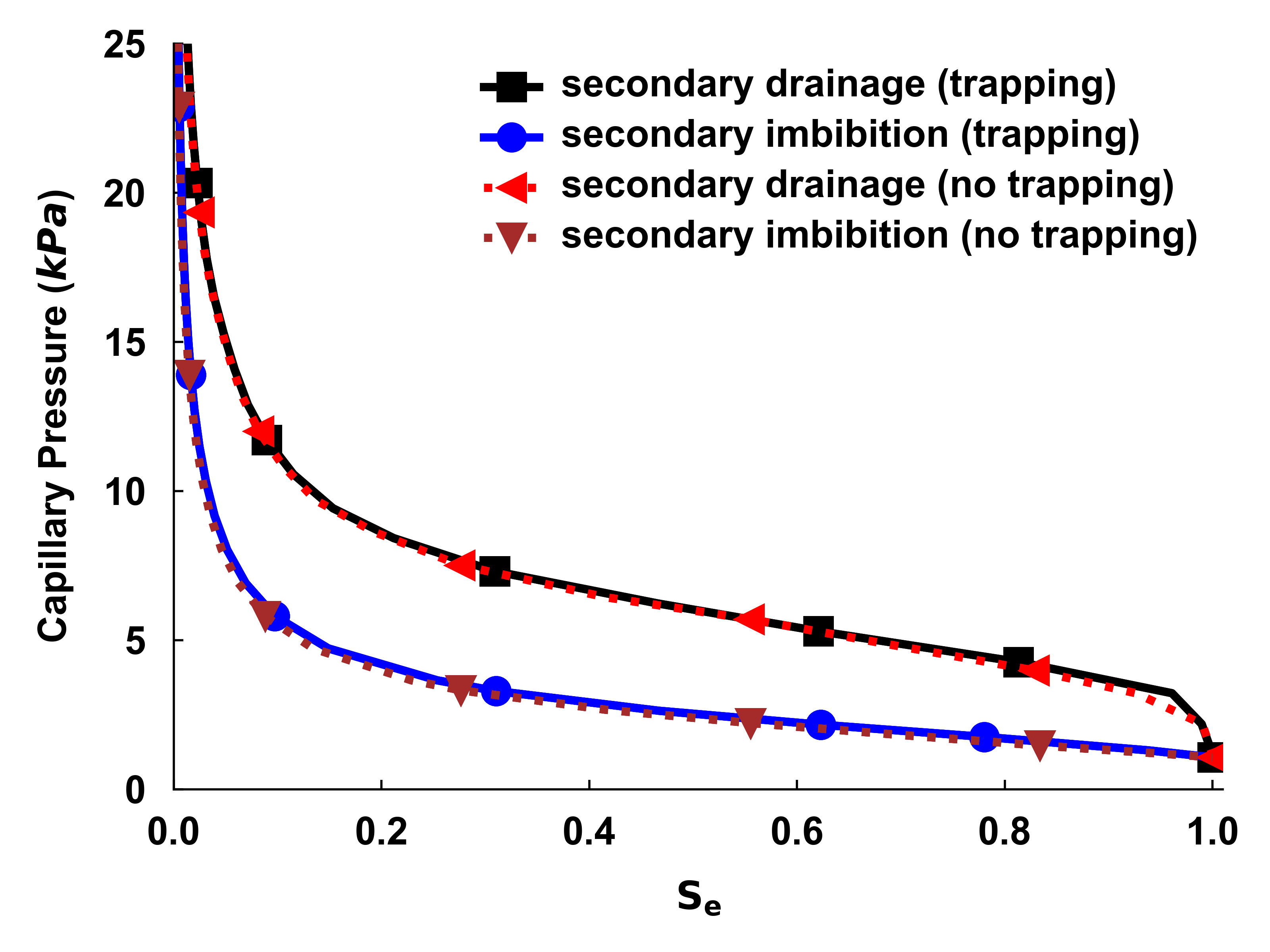}
        \label{PcSe}
        }
    \end{subfigure}
    \begin{subfigure}{
        \includegraphics[width=0.39\textwidth]{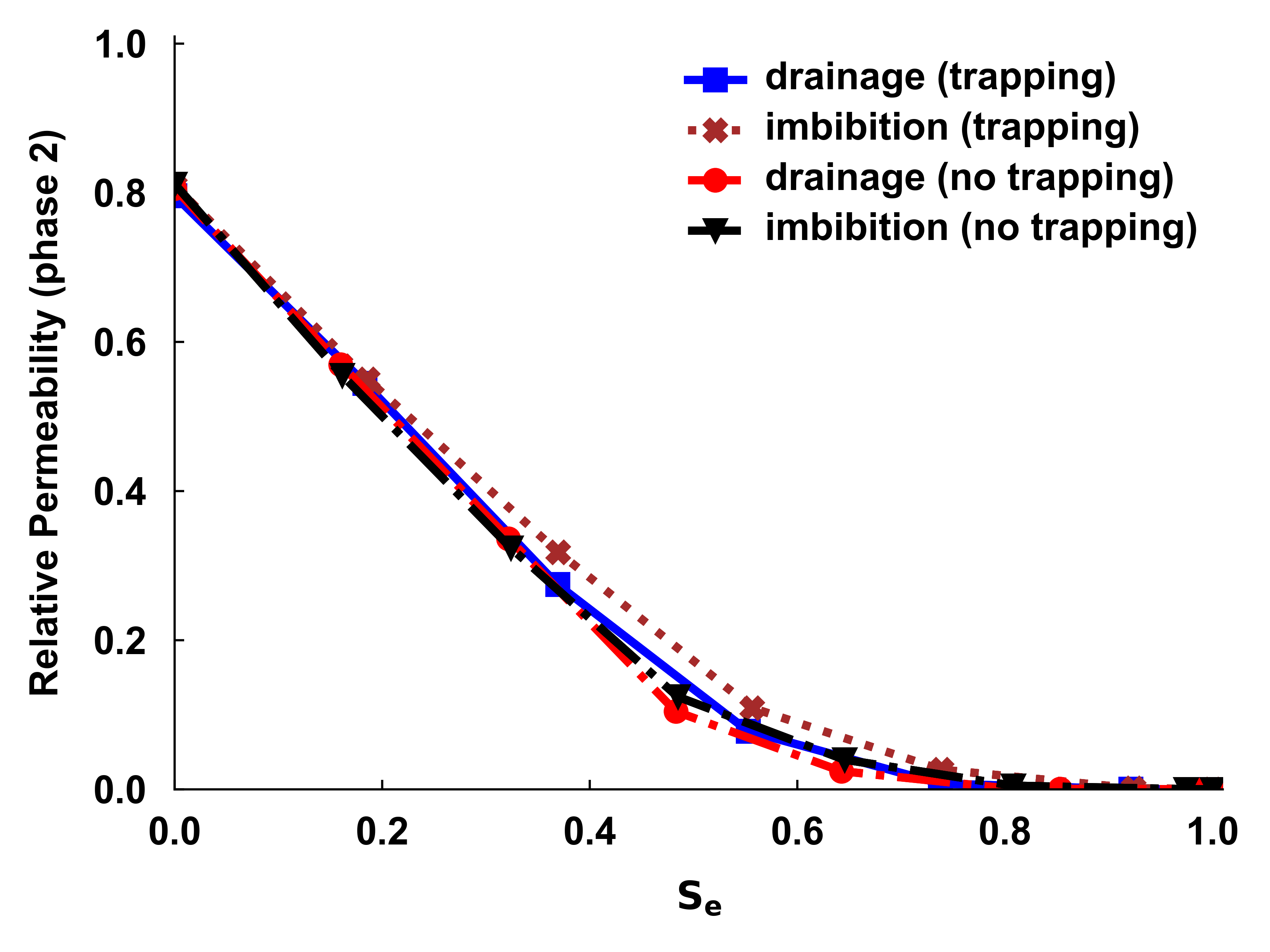}
        \label{kr1Se}
        }
    \end{subfigure}
    \begin{subfigure}{
        \includegraphics[width=0.39\textwidth]{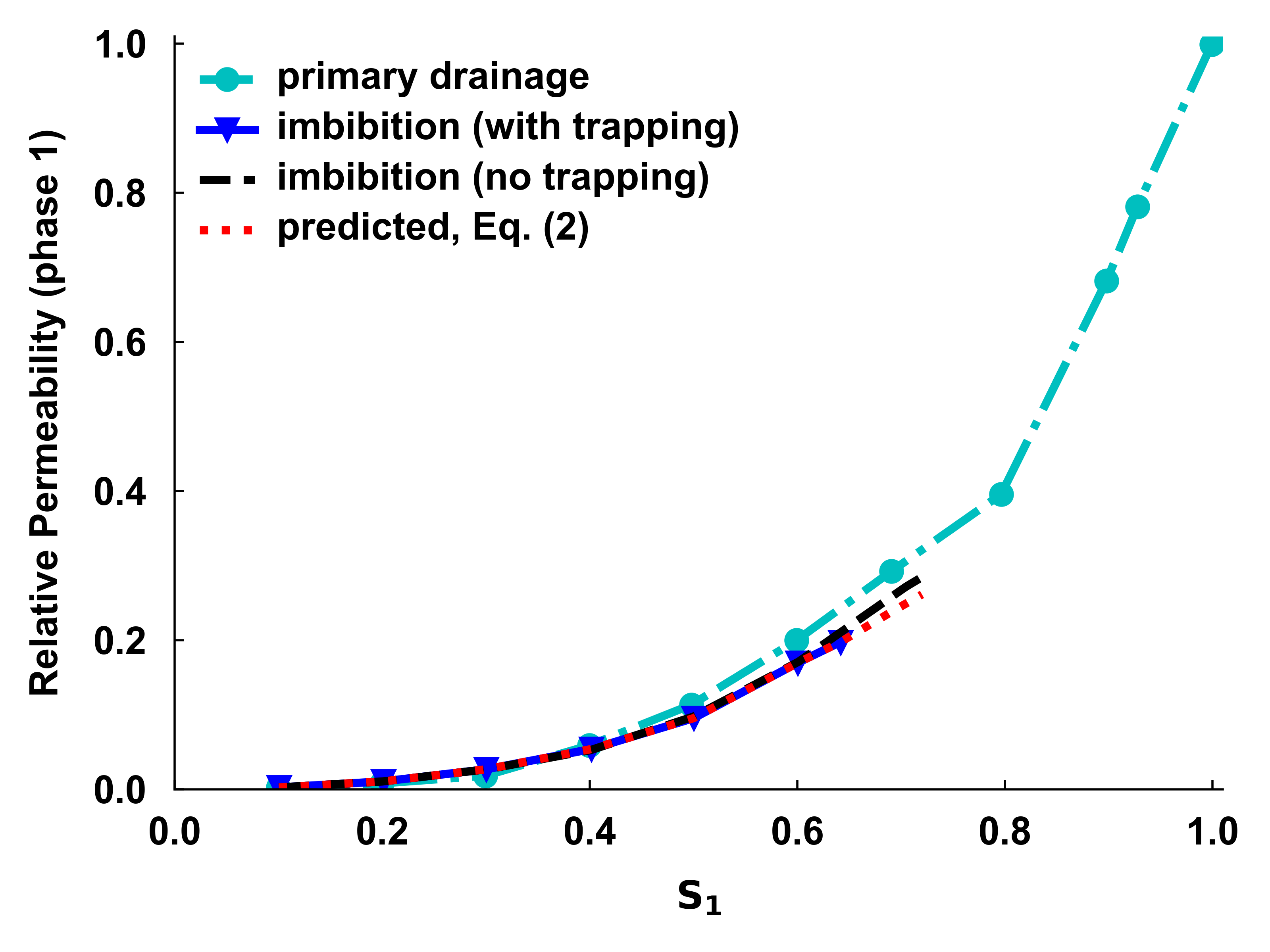}
        \label{kr1S1}
        }
    \end{subfigure}
    
    \caption{(a) Capillary pressure and (b) the relative permeability of phase 2, as a function of the normalized saturation, Eq.~(\ref{Eq:Se}), and (c) phase 1 relative permeability as a function of its own saturation.  For both capillary pressure and the relative permeability of phase 2, the results with and without trapping are similar.  With a rescaling of saturation, measurements with trapping can be converted to predict the behaviour under truly equilibrium conditions. For the relative permeability of phase 2, excellent predictions are also made using the scaling proposed in Eq.~(\ref{Eq:kr}).}
    \label{PckrSeSw}
\end{figure}

We hypothesize that the capillary pressures for displacement cycles after primary drainage with and without trapping are similar when plotted as a function of $S_e$ with appropriate (and different) values for $S_{2t}$.  This is tested in Fig.~\ref{PckrSeSw} which demonstrates that this concept does provide an accurate representation of the capillary pressure hysteresis.

For the wetting phase relative permeability, a larger value is reached without trapping, since the saturation is higher at the end of imbibition.  Here the approach is as follows:  for $S_1 \leq 1-S_{2t}^t$; $k_{r1}^n(S_1) = k_{r1}^t(S_1)$ while for $1-S_{2n}^t\geq S_1 > 1- S_{2t}^t$; we propose:
\begin{equation}
\label{Eq:kr}
    k_{r1}^n(S_1) = k_{r1}^D(S_1) \cdot \frac{k_{r1}^t(S_{1t}^t)}{k_{r1}^D(S_{1t}^t)}
\end{equation}
where the superscript $t$ refers to the case with no Ostwald ripening (trapping), $n$ with Ostwald ripening (the no-trapping model), and $D$ refers to primary drainage.

Fig.~\ref{PckrSeSw} suggests that the scaling proposed in Eq.~(\ref{Eq:kr}) gives a good prediction of the wetting phase relative permeability including Ostwald ripening.  For the other networks studied, see Table~1, our proposed scaling of capillary pressure and relative permeability also works well.

The remaining parameter required to convert measurements without Ostwald ripening is the change in trapped saturation.  This can be found experimentally, by waiting at the end of imbibition for equilibrium to be reached, followed by further injection of phase 1 to allow any connected phase 2 to be displaced.

In the absence of experimental evidence, Table~\ref{tab1} shows the change in trapped saturation for the cases we have studied: Bentheimer sandstone, Estaillades limestone and a reservoir sample.  We see that the reduction in trapped saturation accounting for Ostwald ripening is approximately $20\%$ and this factor could be used to find an initial estimate of $S_{2t}^n$.

\begin{table*}[!]
    \centering
    \caption{Comparison of network properties and predictions of trapped saturations.}
    \setlength{\tabcolsep}{14pt}
    \begin{tabular}{l|c|c|c|c|c}
        \hline
        \hline
        Network & Porosity & $K$ (m$^2$) & $S_{2t}^t$ & $S_{2t}^n$ & $\Delta S$ (\%)\\
        \hline
        Bentheimer  & 0.220 & 2.792 $\times 10^{-12}$ & 0.358 & 0.280 & 21.8 \\
        Estaillades 1 & 0.127 & 1.651 $\times 10^{-13}$ & 0.532 & 0.413 & 22.4\\
        Estaillades 2  & 0.102 & 2.643 $\times 10^{-14}$ & 0.579 & 0.454 & 21.6\\
        Ketton  & 0.074 & 1.108 $\times 10^{-12}$ & 0.393 & 0.300 & 23.7 \\
        Reservoir sample  & 0.133 & 2.339 $\times 10^{-13}$ & 0.464 & 0.366 & 21.1\\
        \hline
        \hline
    \end{tabular}
    \label{tab1}
\end{table*}

\begin{figure}[!]
    \centering
    \includegraphics[width=0.39\textwidth]{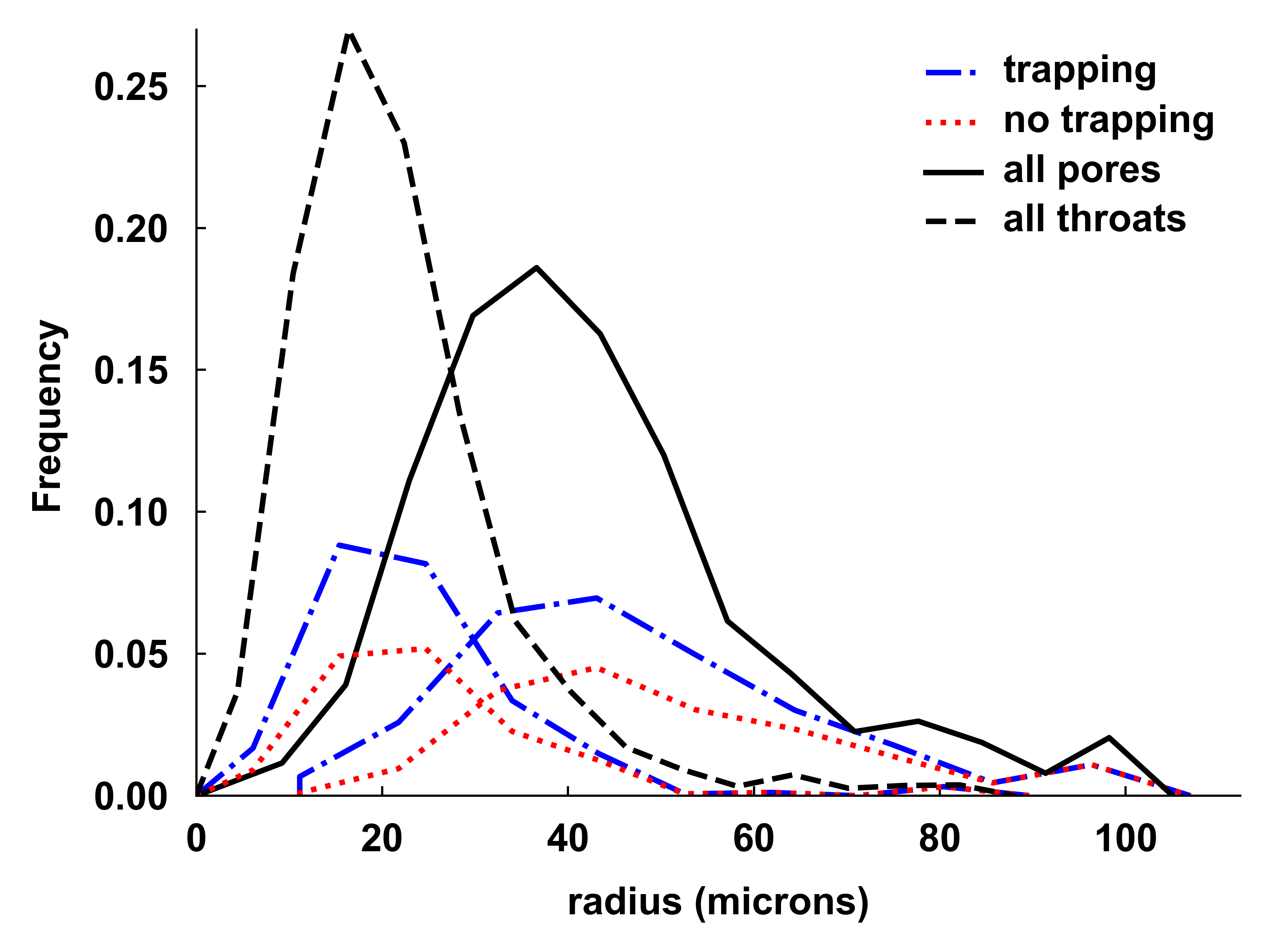}
    \caption{A plot of frequency distribution for the fraction of pores and throats of different radius filled with the phase 2 at the end of imbibition.  Without trapping (no Ostwald ripening) there is less filling overall, but the larger elements are occupied by phase 2.  The difference in the distributions with and without considering Ostwald ripening controls the change in trapped saturation. }
    \label{freq}.
\end{figure}

Fig.~\ref{freq} illustrates the fraction of pores and throats of different radius filled at the end of imbibition for both models.  Allowing for Ostwald ripening leads to a more strictly percolation-like process with a sharp transition from small throats, filled with phase 1, and larger throats occupied by phase 2.  When trapping is accommodated, some smaller throats can remain filled with phase 1, leading to a less steep transition in occupancy.  The resultant difference in trapped saturation will be a function of contact angle (wettability) and pore structure and hence is difficult to predict {\it a priori}.

\section{CONCLUSIONS}
In summary, two-phase displacement where transport of dissolved species allows local capillary equilibrium is a percolation process with trapping, overturning current conventional models \cite{Lenormand1983}.  Experimental measurements on gas/water systems that have not yet reached equilibrium -- which is likely to take days to months \cite{blunt2022ostwald} -- over-estimate the amount of residual trapping by approximately 20\% and the extent of capillary pressure hysteresis.  This is significant for assessments of geological carbon dioxide storage, for instance, where capillary trapping allows secure long-term storage, but whose impact, if based on laboratory experiments, may have been exaggerated \cite{Benson2008}; on the other hand, the reduction in trapping with Ostwald ripening is favourable for hydrogen and natural gas storage and retrieval.

Future work could extend these predictions to systems of arbitrary wettability and to provide direct experimental tests of the changes in multiphase flow properties predicted here. In addition, we could employ a time-dependent pore-scale simulation, as presented by others \cite{deChalendar2018, mehmani2022pore, singh2022level, singh2023pore, singh2024ostwald, moghadasi2023pore} to quantify more precisely the effect of Ostwald ripening during experiments and at representative subsurface conditions.

\section{DATA}
The code used to generate the results in this paper, together with associated data can be found in \href{https://github.com/ImperialCollegeLondon/porescale}{github.com/ImperialCollegeLondon/porescale/OstRipening}.

\section{ACKNOWLEDGEMENT}
AIA would like to thank the Petroleum Technology Development Fund (Nigeria) for funding his PhD.

\bibliography{percolation_without_trapping.bib}

\end{document}